\documentclass[epjST]{svjour}
\usepackage{graphicx}
\usepackage{amsmath}
\usepackage{mathtools}
\usepackage{amssymb}
\usepackage{bm}
\usepackage{pifont}
\usepackage{color}
\usepackage[FIGTOPCAP]{subfigure}
\usepackage[latin1]{inputenc}

\begin{document}
%
%
%
%

\title{
Reply to comment on ``Towards a quantitative kinetic theory of polar active matter'' by Bertin {\em et al.}}

\author{Thomas Ihle\inst{1}\fnmsep\thanks{\email{thomas.ihle@ndsu.edu}}} 

\institute{Department of Physics, North Dakota State University, Fargo, ND 58108-6050, USA}

\abstract{
A reply on the comment of Bertin, Chat{\'e}, Ginelli, Gr{\'e}goire, L{\'e}onard and Peshkov
\cite{ihle_comment_BGL_ST} in this special issue.
}

\maketitle

I am grateful to Bertin {\em et al.} for their valuable comments on Ref. \cite{ihle_ST}.
Since I already discussed the advantages of quantitative versus qualitative modeling in 
Ref. \cite{comment_peshkov_ST}, I will focus on other aspects of their contribution.  
I agree with the authors that, in general, direct tests of kinetic theories are difficult.
Such tests usually require the numerical solution of kinetic equations.
However, at the level of the one-particle density as in Boltzmann- or Enskog-like equations, 
there is no major difficulty.
For two-dimensional active matter, equations of this kind were solved in Refs. {\cite{thueroff_13,ihle_13}.
For example, the kinetic theory results in \cite{ihle_13}
show excellent quantitative agreement with the underlying microscopic
model.
The authors seem to imply that hydrodynamic theories are easier to test.
Concerning simple equations like the Navier-Stokes equation of regular fluids, this might be true,
even though there are good reasons why kinetic descriptions such as the Lattice-Boltzmann method \cite{benzi_92,he_97} 
became quite popular 
for solving hydrodynamic problems. 
For the case of active matter, especially for equations with many nonlinear gradient terms such as Eq. (18) 
in Ref. \cite{ihle_ST},
resolving those terms without creating numerical instabilities could actually be more tedious 
than directly dealing with a kinetic equation.

I mostly agree with the authors about the role of fluctuations on the jump of the order parameter, which is discussed in the caption of their Fig. 1.
However, when debating how accurately the transition line of a discontinuous transition can be predicted by a theory,
one should keep in mind that the hysteresis region is quite small. For example, its size is about $14 \%$ 
of the critical $\sigma$ in Fig. 1  
in \cite{ihle_comment_BGL_ST}. Thus, as long as the linear stability threshold and the maximum $\sigma$ for the downward jump
of the order parameter 
are accurately predicted, theoretical estimates for the exact transition line will only have an error bar of a few percent.

I appreciate that the authors bring up the different scaling of the density with 
the expansion parameter $\epsilon$, something I also discussed in \cite{comment_peshkov_ST}.
I agree that if the most recent scaling of Bertin {\em et al.}'s approach, denoted as BDG, from Ref. \cite{peshkov_ST} 
is applied to the hydrodynamic equation Eq. (18) in \cite{ihle_ST},
at order $O(\epsilon^3)$ the same structure of a Navier-Stokes-like equation 
as in their work would appear. 
However, the transport coefficients would necessarily be different due to the different collision frequencies, 
collisional momentum transfer (absent in Boltzmann-approaches) and the discrete time step, see discussion in
\cite{ihle_ST,foot1}.
When looking back at the original papers of the authors, \cite{bertin_06,bertin_09}, I noticed that
they used to have the same scaling of the zeroth Fourier coefficient and the density, $f_0\sim O(1)$, $\rho\sim O(1)$,
that is applied in the phase space approach (PSA) \cite{ihle_ST,ihle_11}. This can be seen in Eq. (24) of \cite{bertin_09}
and leads to several terms in the equation of the momentum density, Eq. (27) in Ref. \cite{bertin_09}, 
which are now classified as higher order, $O(\epsilon^4)$,
acording to the newest scaling ansatz of the authors. 
This might have contributed to the confusion whether the density gradient terms in PSA that are absent in BDG
should be labeled as higher order or not.

I was surprised by the author's clarification of the binary aligment interactions that are supposed to occur only once per collision, exactly when two particles reach interaction range for the first time.
To my knowledge, this has never been mentioned in the foundation papers on BDG such as Refs. \cite{bertin_06,bertin_09}
but also not in the newest review, Ref. \cite{peshkov_ST}. 
According to this interaction rule, 
two particles would be invisible to each other once they are inside each others collision sphere after the first interaction. 
This would be quite similar to the high speed regime of the Vicsek-model 
where particles only have a chance to ``see'' each 
other at the end of the streaming step.
If this one-alignment-only rule of \cite{ihle_comment_BGL_ST} 
is indeed the microscopic model behind the theory of Bertin {\em et al.} and if according to their claims 
a Boltzmann description is adequate, it should be easy for them to check agreement between 
their theory and agent-based simulations. 
Thus,  it is a bit confusing to me that in Refs. \cite{bertin_06,bertin_09} the authors used the Vicsek-model 
in their agent-based comparisons 
instead of this different 
but well-defined microscopic model.

Finally, referring to extreme limits of the Vicsek model, the authors write that they are not interested
in the limits of zero and infinite speed, and that these limits are singular. 
I also believe that the zero speed limit is singular but I am quite sure 
that the other limit of infinite speed is rather well behaved in the sense that an expansion around this limit
in the parameter $\varepsilon=R/(v_0 \tau)$ is not singular \cite{chou_14}. 
This view is supported by the fact that the phase diagram obtained analytically 
at infinite speed agrees well
with agent-based simulations of the Vicsek-model at moderate speeds, at least at larger 
densities, $M=\rho \pi R^2>1$, 
see Fig. (1) of Ref. \cite{ihle_11}.
I agree that the two limits per se are not that interesting but if a theory makes sense in two extreme cases,
the likelihood is larger that it is also ok inbetween these limits.


\begin{thebibliography}{99}
\bibitem{foot1}
Note, the transport coefficients derived in \cite{ihle_11} do not include contributions from
collisional momentum transfer because they are negligible anyways in the considered limit
of large mean free path, $v_0\tau/R\gg 1$. For details on how to calculate them, see Ref. \cite{ihle_09}.
\end{thebibliography}
\end{document}